\documentclass[aps,prb,reprint,superscriptaddress]{revtex4-1}
\usepackage{amsfonts}
\usepackage{amssymb}
\usepackage{amsmath}
\usepackage{graphicx}
\usepackage{epsfig}

\begin{document}

\title{Revealing the non-adiabatic and non-Abelian multiple-band effects via
anisotropic valley Hall conduction in bilayer graphene}
\author{Ci Li}
\email{oldsmith@hku.hk}
\affiliation{Department of Physics, The University of Hong Kong, Hong Kong,
China}
\affiliation{HKU-UCAS Joint Institute of Theoretical and
Computational Physics at Hong Kong, China}
\author{Matisse Wei-Yuan Tu}
\email{kerustemiro@gmail.com}
\affiliation{Department of Physics, National Sun Yat-sen University, Taiwan}
\author{Wang Yao}
\affiliation{Department of Physics, The University of Hong Kong, Hong Kong,
China}
\affiliation{HKU-UCAS Joint Institute of Theoretical and
Computational Physics at Hong Kong, China}

\begin{abstract}
Many quantum materials of interest, ex., bilayer graphene, possess a number
of closely spaced but not fully degenerate bands near the Fermi level, where
the coupling to the far detuned remote bands can induce Berry curvatures of
the non-Abelian character in this active multiple-band manifold for
transport effects. Under finite electric fields, non-adiabatic interband
transition processes are expected to play significant roles in the
associated Hall conduction. Here through an exemplified study on the valley
Hall conduction in AB-stacked bilayer graphene, we show that the
contribution arising from non-adiabatic transitions around the bands near
the Fermi energy to the Hall current is not only quantitatively about an
order-of-magnitude larger than the contribution due to adiabatic
inter-manifold transition with the non-Abelian Berry curvatures. Due to the
trigonal warping, the former also displays an anisotropic response to the
orientation of the applied electric field that is qualitatively distinct
from that of the latter. We further show that these anisotropic responses
also reveal the essential differences between the diagonal and off-diagonal
elements of the non-Abelian Berry curvature matrix in terms of their the
contributions to the Hall currents. We provide a physically intuitive
understanding on the origin of distinct anisotropic features from different
Hall current contributions, in terms of band occupations and interband
coherence. This then points to the generalization beyond the specific
example of bilayer graphenes.
\end{abstract}

\maketitle

\section{Introduction}

Band structure effects on transport of electrons driven by an external
electric field constitute one of the most fundamental issues in solid-state
physics \cite{Mer}. In principle, one can group the electronic bands into
two kinds of manifolds, namely, the active and the remote, according to the
relation between the relevant gaps and the applied field. The active
manifold contains those bands around the Fermi energy which play the
dominant roles in the field-driven transport. Other bands are categorised
into the remote manifold. Due to the relatively large gaps between the
active and the remote manifolds, the field-induced inter-manifold
transitions are well captured by the adiabatic approximation, in which the
electric field is treated as a perturbation. When the active manifold
contains a single band or a number of fully degenerate bands, this adiabatic
description has been successful in manifesting one of the most non-trivial
band structure effects on electron transport, namely, the Hall effect \cite%
{Kar,Lutt,Adam,Thou,Mac,Naga,Niu}, which includes a number of varieties such
as spin \cite{Zhang,Murakami04235206,Mac1,Kane} and valley Hall effects \cite%
{Niu1,Yao}. On the other hand, when the active manifold contains a number of
closely spaced but not fully degenerate bands, the effects of a finite
electric field on the interband transitions within the active manifold are
beyond the validity of perturbation treatment and therefore also requires
non-adiabatic consideration. Correspondingly, this leads to a non-adiabatic
Hall effect, which has been found to be pronounced in Dirac cones with a
Berry curvature hotspot, as recently discussed \cite{Tu1} and extended to
situations with spatially varied band structures \cite{Tu}.

Although previous studies have respectively expounded on the Hall effects in
the adiabatic regime and expanded into the non-adiabatic one, interesting
realistic materials such as transition metal dichalcogenides (ex. twisted $%
\text{MoSe}_{2}$ homobilayers \cite{HY,Mac2,DW,HY1}) and bilayer graphenes%
\cite{Mc2} indeed possess band structures that demand a coherent unification
of the above two perspectives. Explicitly, from the view of the carrier
motion, the Hall effect is rooted to the anomalous part of the velocity. On
one end, when the active manifold contains only a single band or fully
degenerate bands over the whole Brillouin zone, the adiabatic inter-manifold
transition results in the well-known expression of the anomalous velocity in
terms of the Berry curvatures of the active bands \cite%
{Thou,Niu,Tu,NiuB,Boehm03book}. On the other end, when the remote bands are
completely ignored, non-adiabatic dynamics among the non-degenerate bands
within the active manifold gives rise to a renormalized carrier velocity due
to the finite electric field, that also has an anomalous part contributing
to the corresponding non-adiabatic Hall effect \cite{Tu1}. Naturally, the
consequence from the interplay of these two ends is an important question
worthy further investigation, since many quantum materials of interest do
have non-degenerate active bands together with non-negligible coupling to
remote bands.

Apart from the non-adiabatic effects mentioned above, the multiple-band
nature of the active manifold, as well-known, also gives rise to the
non-Abelian characters of Berry curvatures upon projecting out the remote
bands \cite{Niu} (see also \cite{Tu} for pedagogical derivation). On one
hand, from the current response to an infinitesimal applied field, the pure
geometrical aspects of the non-Abelian characters on the Hall currents have
been well understood \cite{Murakami04235206,Shi}. On the other hand, the
non-Abelian Berry curvatures as anomalous driving forces for the dynamics of
single wave packets have also been analysed through the peculiarly induced
motion under various contexts \cite%
{Culcer05085110,Chang08193202,Gorbar18045203,Stedman19103007}. Nevertheless,
the interplay between these two faces of multiple-band effects, namely, the
non-Abelian characters and the non-adiabatic dynamics still remains an open
question.

\begin{figure}[tbp]
\begin{center}
\includegraphics[width=0.48\textwidth]{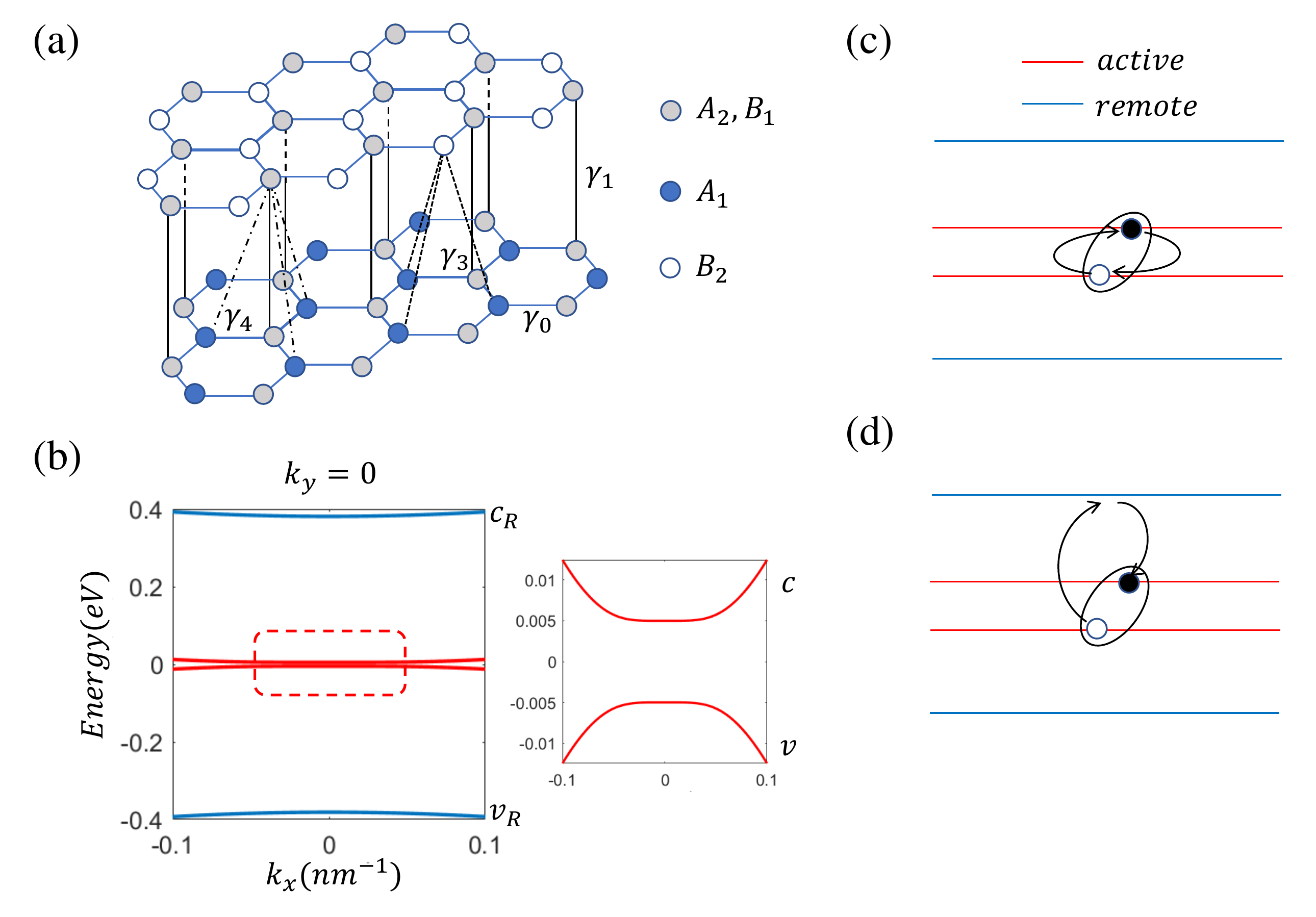}
\end{center}
\caption{(Color online) (a): The atomic structure of an $AB$-stacked bilayer
graphene in side view. (b): The four-band dispersion (with $\protect\gamma %
_{3}=0$ in (a), see also Eq.~(\protect\ref{H})) clearly distinguishing the
active and the remote bands. (c): Interband coherence directly formed
through non-adiabatic intra-active-manifold dynamics. (d): Interband
coherence indirectly formed through adiabatic inter-manifold transitions.
They imply distinct roles in their respective contributions to the Hall
conduction.}
\label{fig1}
\end{figure}

In this work, we address the above raised issues through investigating the
valley Hall currents in biased graphene bilayer with a modest gap ($0.01%
\mathrm{eV}$) (Fig.~\ref{fig1}(a)) \cite%
{Novo,Mc,Ohta,Mc1,Cast,Oost,YB,Yang,Mc2,Ju}, The effective four-band model
for this material features two bands close in energy forming the active
manifold and two other remote bands separated from the active bands by
around 0.4eV (see Fig. 1(b)), suitable for the present purpose.

Explicitly, we investigate the transverse valley currents of $AB$-stacked
bilayer graphene under finite external electric fields. We treat both the
contributions from the adiabatic inter-manifold-transitions and
non-adiabatic intra-active-manifold transitions to the Hall conduction on an
equal footing. The main point of the present research is to establish that
the non-adiabatic aspect of the Hall current is a distinct facet that is
qualitatively different from the more familiar Hall effect usually studied
in the adiabatic regime. This is revealed through the dependence of the Hall
current on the angle between the applied electric field and the crystalline
axis, exploiting the anisotropy from trigonal warping of the electronic band
structure of AB-stacked bilayer graphene. The ability to apply an electric
field along various directions across a sample has been experimentally
realized \cite{Kang19324} with an attempt to facilitate the study of the
well-known nonlinear Hall effect supported by \textit{Abelian} Berry
curvature dipoles \cite{note-i-0}. We show that the feasibility on the
angular-dependence of the Hall current not only reveals the crucial
distinctions between the adiabatic inter-manifold and the non-adiabatic
intra-active-manifold contributions to the Hall current. But it also
uncovers interesting difference between contributions involving diagonals
and off-diagonals from the \textit{non-Abelian} Berry curvatures matrix.

%The adiabatic inter-manifold contribution to the Hall current in the
%Mechanistically speaking,

%The adiabatic aspect is fully characterised by the diagonals of the non-Abelian Berry curvature matrix comprising the linear conductivity. The non-adiabatic aspect, however, is manifested through both the off-diagonals of the non-Abelian Berry curvature matrix and anisotropic response of the Hall current to the applied electric field. The anisotropic response has been utilised to characterise the nonlinear-in-electric-field effects in experiments \cite{Kang19324} for probing the so-called nonlinear Hall effect arising from $ac$-driven extrinsic .

\vspace{-0.06cm}

\section{Unified description of adiabatic and non-adiabatic regimes}

To begin, we first summarise the main framework that unify the wave-packet
description of electronic transport for both the adiabatic and the
non-adiabatic regimes \cite{Tu}. We consider a general band structure,
described by the Hamiltonian, $\mathcal{H}\left( \boldsymbol{k}\right) $
that totally has $N_{A}$ bands in the active manifold $A$ and $N_{R}$ bands
in the remote manifold $R$. The band wavefunctions and energies of $\mathcal{%
H}\left( \boldsymbol{k}\right) $ are denoted respectively by $\left\vert
u_{n}\left( \boldsymbol{k}\right) \right\rangle $ and $\varepsilon
_{n}\left( \boldsymbol{k}\right) $ with the band index $n$ and the Bloch
wave vector $\boldsymbol{k}$. The separation between the manifolds $A$ and $%
R $ means $\forall n_{a},m_{a}\in A$ and $\forall n_{r}\in R$ that
\begin{equation*}
\left\vert \varepsilon _{n_{a}}-\varepsilon _{n_{r}}\right\vert \gg \underset%
{m\in {A}}{\text{MAX}}\left\vert \varepsilon _{n_{a}}-\varepsilon
_{m}\right\vert ,~\frac{\left\vert \varepsilon _{n_{a}}-\varepsilon
_{n_{r}}\right\vert }{\left\vert \varepsilon _{m_{a}}-\varepsilon
_{n_{r}}\right\vert }\approx 1,
\end{equation*}%
The dynamics of a single wave packet is described by the time dependent Schr%
\"{o}dinger equation $i\hbar \left\vert \dot{u}\left( t\right) \right\rangle
=\mathcal{H}\left( \boldsymbol{k}\right) \left\vert u\left( t\right)
\right\rangle $ in which the carrier's Bloch wave vector acquires its time
dependence $\hbar \boldsymbol{\dot{k}}=-e\boldsymbol{E}$ through the driving
of an external electric field $\boldsymbol{E}$. The solution has the form $%
\left\vert u\left( t\right) \right\rangle =\sum_{n\in {A}+R}\eta _{n}\left(
t\right) e^{i\gamma _{n}\left( \boldsymbol{k}\right) }\left\vert u_{n}\left(
\boldsymbol{k}\right) \right\rangle $ with the band amplitudes $\eta _{n}$
subject to normalisation. Here $\gamma _{n}\left( \boldsymbol{k}\right)
=\int_{\boldsymbol{k}_{0}}^{\boldsymbol{k}}\text{d}\boldsymbol{k}^{\prime
}\cdot \left[ \mathcal{R}_{\boldsymbol{k}^{\prime }}\right] _{n,n}$ is the
Berry phase of band $n$ for a trajectory of $\boldsymbol{k}$ starting from $%
\boldsymbol{k}_{0}$ and $\left[ \mathcal{R}_{\boldsymbol{k}}\right]
_{n,m}=i\left\langle u_{n}\right\vert \frac{\partial }{\partial \boldsymbol{k%
}}\left\vert u_{m}\right\rangle $ denotes the Berry connection between bands
$n$ and $m$.

By projecting out the remote bands while retaining their effects, the
centre-of-mass motion of a single wave packet is described by the following
equations,
\begin{subequations}
\label{eom}
\begin{align}
\dot{\boldsymbol{r}}& =\left\langle \left[ \mathcal{D}_{\hbar \boldsymbol{k}%
},\mathcal{H}_{A}\left( \boldsymbol{k}\right) \right] -\dot{\boldsymbol{k}}%
\times \boldsymbol{\mathcal{F}}_{\boldsymbol{k}}\right\rangle _{A},
\label{eom-cc} \\
\hbar \dot{\boldsymbol{k}}& =-e\boldsymbol{E}
\end{align}%
where $\mathcal{H}_{A}\left( \boldsymbol{k}\right) $ denotes the projection
of $\mathcal{H}\left( \boldsymbol{k}\right) $ into the active manifold and
\begin{equation}
\left[ \mathcal{D}_{\boldsymbol{k}}\right] _{n,m}=\delta _{n,m}\frac{
\partial }{\partial \boldsymbol{k}}-i\left[ \mathcal{R}_{\boldsymbol{k}} %
\right] _{n,m},
\end{equation}
is the covariant derivative. The boldface $\boldsymbol{\mathcal{F}}_{%
\boldsymbol{k}}$ stands for a rank-1 tensor as a vector whose spatial
components $\mathcal{F}_{\boldsymbol{k}}^{\lambda }$'s are related to the
rank-2 Berry curvature tensor $(1/2)\epsilon ^{\alpha \beta \lambda }\left[
\mathcal{F}_{\boldsymbol{k}}^{{\alpha }{\beta }}\right] =\mathcal{F}_{%
\boldsymbol{k}}^{\lambda }$ where $\epsilon ^{\alpha \beta \lambda }$ is the
Levi-Civita symbol and
\begin{equation}
\mathcal{F}_{\boldsymbol{k}}^{\alpha \beta }=\left\{ \frac{\partial \left[
\mathcal{R}_{k_{\beta }}\right] }{\partial {k}_{\alpha }}-\frac{\partial %
\left[ \mathcal{R}_{k_{\alpha }}\right] }{\partial {k}_{\beta }}-i\left[
\mathcal{R}_{k_{\alpha }},\mathcal{R}_{k_{\beta }}\right] \right\} ,
\label{NAB}
\end{equation}
which retains the non-Abelian matrix structure of the Berry connections $%
\mathcal{R}_{\boldsymbol{k}}$ in band indices. Here the square bracket
stands for the average $\left\langle {O}\right\rangle _{A}=\sum_{n,m\in {A}%
}\eta _{n}^{\ast }O_{n,m}\eta _{m}$ over the band amplitudes $\eta _{n}$'s,
which are subject to
\end{subequations}
\begin{subequations}
\label{movH}
\begin{equation}
i\hbar \dot{\boldsymbol{\eta }}=\bar{\mathcal{H}}\boldsymbol{\eta }
\end{equation}%
where the matrix elements of the moving-frame Hamiltonian $\bar{\mathcal{H}}$
read
\begin{equation}
\bar{\mathcal{H}}_{n,m}=\delta _{n,m}\varepsilon _{n}+\left( 1-\delta
_{n,m}\right) e\boldsymbol{E}\cdot \left[\bar{\mathcal{R}}_{\boldsymbol{k}}%
\right]_{n,m},  \label{movH-2}
\end{equation}
in which $\left[ \bar{\mathcal{R}}_{\boldsymbol{k}}\right]
_{n,m}=e^{-i\gamma _{n}}\left[ \mathcal{R}_{\boldsymbol{k}}\right]
_{n,m}e^{i\gamma _{m}}$. The non-Abelian Berry curvature $\mathcal{F}_{%
\boldsymbol{k}}^{\alpha \beta }$ given by Eq.~(\ref{NAB}) arises from
adiabatic inter-manifold transitions and is independent of $\boldsymbol{E}$.
Nevertheless, its influences on the carrier velocity $\dot{\boldsymbol{r}}$,
manifested through the second term in Eq.~(\ref{eom-cc}), namely, $%
\left\langle \boldsymbol{\mathcal{F}}_{\boldsymbol{k}}\right\rangle _{A}$,
contains the non-adiabatic intra-active-manifold dynamics embedded in $%
\boldsymbol{\eta}$ that displays non-perturbation effects from a finite $%
\boldsymbol{E}$ (see Eq.~(\ref{movH})).

The electrical currents arise from an ensemble of wave packets. By taking
into accounts the scattering effects and the decoherence within the band
space \cite{Tu}, the electric current is found to be
\end{subequations}
\begin{subequations}
\label{crnt}
\begin{equation}
\boldsymbol{J}=-e\sum_{i}\int_{\boldsymbol{k}}g_{i}\left( \boldsymbol{k}%
\right) \bar{\boldsymbol{v}}_{i}\left( \boldsymbol{k}\right),
\label{Bltzeq-4}
\end{equation}%
where we have abbreviated $\int \text{d}\boldsymbol{k}$ by $\int_{%
\boldsymbol{k}}$. Here $\bar{\boldsymbol{v}}_{i}$ is the ensemble-averaged
velocity, containing two contributions,
\begin{equation}
\bar{\boldsymbol{v}}_{i}=\bar{\boldsymbol{v}}_{i}^{A}+\bar{\boldsymbol{v}}%
_{i}^{R}.  \label{eqvelo-x-i}
\end{equation}%
The intra-active-manifold non-adiabatic dynamics contributes with
\end{subequations}
\begin{subequations}
\label{crnt-A}
\begin{equation}
\bar{\boldsymbol{v}}_{i}^{A}=\bar{\boldsymbol{v}}_{i}^{\text{occ},A}+\bar{%
\boldsymbol{v}}_{i}^{\text{coh},A}  \label{decVelo-A}
\end{equation}%
where
\begin{equation}
\bar{\boldsymbol{v}}_{i}^{\text{occ},A}=\sum_{n\in {A}}\left\vert \bar{\eta}%
_{n}^{i}\right\vert ^{2}\frac{\partial \varepsilon _{n}}{\partial
\boldsymbol{k}},  \label{2x2-b}
\end{equation}%
is the normal velocity due to band dispersion and
\begin{equation}
\bar{\boldsymbol{v}}_{i}^{\text{coh},A}=\sum_{n\neq {m}\in {A}}\left( \bar{%
\eta}_{n}^{i}\right) ^{\ast }\left[ \frac{\partial \mathcal{H}_{A}}{\partial
\hbar \boldsymbol{k}}\right] _{n,m}\bar{\eta}_{m}^{i},  \label{2x2-A}
\end{equation}%
is one part of the anomalous velocity. The other part of the anomalous
velocity is contributed by the inter-manifold adiabatic dynamics with
\end{subequations}
\begin{subequations}
\label{decVelo-R}
\begin{equation}
\bar{\boldsymbol{v}}_{i}^{R}=\bar{\boldsymbol{v}}_{i}^{\text{occ},R}+\bar{%
\boldsymbol{v}}_{i}^{\text{coh},R},  \label{2x2}
\end{equation}%
where
\begin{equation}
\bar{\boldsymbol{v}}_{i}^{\text{occ},R}=\frac{e}{\hbar }\boldsymbol{E}\times
\sum_{n\in {A}}\left\vert \bar{\eta}_{n}^{i}\right\vert ^{2}\left[
\boldsymbol{\mathcal{F}}_{\boldsymbol{k}}\right] _{n,n},  \label{2x2-Rocc}
\end{equation}%
and
\begin{equation}
\bar{\boldsymbol{v}}_{i}^{\text{coh},R}=\frac{e}{\hbar }\boldsymbol{E}\times
\sum_{n\neq m\in {A}}\left( \bar{\eta}_{n}^{i}\right) ^{\ast }\left[
\boldsymbol{\mathcal{F}}_{\boldsymbol{k}}\right] _{n,m}\bar{\eta}_{m}^{i}.
\label{2x2-Rcoh}
\end{equation}%
come from projecting the inter-manifold dynamics onto the band occupations $%
\left\vert \bar{\eta}_{n}^{i}\right\vert ^{2}$ and interband coherence $%
\left( \bar{\eta}_{n}^{i}\right) ^{\ast }\bar{\eta}_{m}^{i}$ in the active
manifold respectively. Here $\bar{\boldsymbol{\eta }}^{i}$, a vector with
components $\bar{\eta}_{n}^{i}$ enumerated by $n$, is defined as an
eigenvector of the moving-frame Hamiltonian's projection on active bands, $%
\bar{\mathcal{H}}_{A}\left( \boldsymbol{k},\boldsymbol{E}\right) $, indexed
with $i$, namely, $\bar{\mathcal{H}}_{A}\bar{\boldsymbol{\eta }}^{i}=%p
\mathcal{E}_{i}\bar{\boldsymbol{\eta }}^{i}$ with $\mathcal{E}_{i}$ the
corresponding eigenvalue. We call $\bar{\boldsymbol{\eta }}^{i}$'s the
hybridised bands and they depend on the electric field in a non-perturbation
way \cite{Tu1,Tu}. The physical meaning is the following. The joint action
of decoherence and the electric field favours a certain form of interband
coherence as those contained in $\bar{\boldsymbol{\eta }}^{i}$'s. As a
result, the carriers are led to a statistical mixture of the hybridised
bands $\bar{\boldsymbol{\eta }}^{i}$'s. The distribution function with
respect to occupations on the hybridised bands is given by $%
g_{i}=g_{i}^{0}+\delta g_{i}$ where $g_{i}^{0}=1/\left[ \exp \left( \frac{%
\mathcal{E}_{i}-\mu }{k_{B}T}\right) +1\right] $ with $\mu $ the chemical
potential and $T$ the temperature and $\delta g_{i}=\left( e/\hbar \right)
\tau \boldsymbol{E}\cdot \partial g_{i}^{0}/\partial \boldsymbol{k}$ with $%
\tau $ the relaxation time \cite{note-1}.

The current Eq.~(\ref{crnt}) can be further decomposed as
\end{subequations}
\begin{equation*}
\boldsymbol{J}=\boldsymbol{J}^{\text{L}}+\boldsymbol{J}^{\text{H}},
\end{equation*}%
where $\boldsymbol{J}^{\text{L}}=-e\int_{\boldsymbol{k}}\sum_{i}\delta g_{i}%
\bar{\boldsymbol{v}}_{i}$ and
\begin{subequations}
\label{crnt-dH}
\begin{equation}
\boldsymbol{J}^{\text{H}}=\boldsymbol{J}_{A}^{\text{H}}+\boldsymbol{J}_{R}^{%
\text{H}},
\end{equation}%
in which
\begin{equation}
\boldsymbol{J}_{A}^{\text{H}}=-e\int_{\boldsymbol{k}}\sum_{i}g_{i}^{0}\bar{%
\boldsymbol{v}}_{i}^{\text{coh},A},~\boldsymbol{J}_{R}^{\text{H}}=-e\int_{%
\boldsymbol{k}}\sum_{i}g_{i}^{0}\bar{\boldsymbol{v}}_{i}^{R}
\label{crnt-dH-2}
\end{equation}%
on knowing that $\int_{\boldsymbol{k}}g_{i}^{0}\bar{\boldsymbol{v}}_{i}^{%
\text{occ},A}=0$ \cite{Mer}. These two contributions $\boldsymbol{J}^{\text{L%
}}$ and $\boldsymbol{J}^{\text{H}}$ respectively reproduce the
well-established expressions of the longitudinal current and the Hall
current in the limit of infinitesimal $\boldsymbol{E}$ \cite{Niu,Tu}. In
this work, we are only interested in $\boldsymbol{J}^{\text{H}}$.

Within this framework, the issues raised in the introduction can now be
addressed. The first question is on the interplay between the respective
contributions to the Hall current arising from adiabatic coupling to remote
bands and from non-adiabatic dynamics within the active manifold. This can
be approached from the well-known notion that the Hall current is mediated
by the interband coherence \cite{Naga}. Both contributions $\boldsymbol{J}%
_{A}^{\text{H}}$ and $\boldsymbol{J}_{R}^{\text{H}}$ contain interband
coherence among the active bands, namely, $\left( \bar{\eta}_{n}^{i}\right)
^{\ast }\bar{\eta}_{m}^{i}$ for $n\neq {m}$ with $n,m\in {A}$. Such
coherence in $\boldsymbol{J}_{A}^{\text{H}}$ explicated by Eq.~(\ref{2x2-A})
is directly formed through the field-induced interband transitions within
the active manifold. In contrast, for $\boldsymbol{J}_{R}^{\text{H}}$, such
coherence given by Eq.~(\ref{2x2-Rcoh}) is only indirectly formed through
transitions forth and back between the active and the remote bands (see
illustrations in Fig.~\ref{fig1}(c) and (d) respectively). This leads us to
anticipate that the natures of the Hall current manifested via $\boldsymbol{J%
}_{A}^{\text{H}}$ and $\boldsymbol{J}_{R}^{\text{H}}$ can be very different.

The second question concerns the two faces of the multiple-band effects,
namely, the non-adiabatic dynamics and the non-Abelian characters of the
Berry curvatures. This can be viewed from the decomposition
\end{subequations}
\begin{subequations}
\label{remote-H-decomp}
\begin{equation}
\boldsymbol{J}_{R}^{\text{H}}=\boldsymbol{J}_{\text{occ},R}^{\text{H}}+%
\boldsymbol{J}_{\text{coh},R}^{\text{H}},  \label{remote1}
\end{equation}%
where
\begin{equation}
\boldsymbol{J}_{\text{occ}/\text{coh},R}^{\text{H}}=-e\int_{\boldsymbol{k}%
}\sum_{i}g_{i}^{0}\bar{\boldsymbol{v}}_{i}^{\text{occ}/\text{coh},R}.
\label{remote2}
\end{equation}%
When the active bands are fully degenerate over the Brillouin zone, $%
\boldsymbol{J}_{\text{coh},R}^{\text{H}}$ has no contribution to the Hall
current, as we have discussed in Ref.\cite{Tu}. There we have also shown
that the Hall current reduces to the well-established expression $%
\boldsymbol{J}^{\text{H}}=\left( e^{2}/\hbar \right) \boldsymbol{E}\times
\int_{\boldsymbol{k}}\text{Tr}\left( \boldsymbol{\mathcal{F}}_{\boldsymbol{k}%
}\right) $ \cite{Shi} when the degenerate band energy is set below the Fermi
energy. The gauge symmetry associated with band degeneracy can be utilized
to study the non-Abelian characters \cite{Wilczek842111}. For general
non-degenerate bands, where gauge symmetry is not expected, the non-Abelian
characters of $\mathcal{F}_{\boldsymbol{k}}^{z}$ as matrices are fully
embedded in the decomposition Eq.~(\ref{remote-H-decomp}). In conjunction
with Eq.~(\ref{decVelo-R}), we find the two current contributions $J_{\text{%
occ},R}^{\text{H}}$ and $J_{\text{coh},R}^{\text{H}}$ involve only the
diagonals and the off-diagonals of $\mathcal{F}_{\boldsymbol{k}}^{z}$
respectively, bearing also distinct physical meanings as currents arising
from band occupations and interband coherence. Comparison between $J_{\text{%
occ},R}^{\text{H}}$ and $J_{\text{coh},R}^{\text{H}}$ thus provides a mean
to reveal the multiple-band effects that underlie the Hall currents
including both the non-Abelian characters of $\mathcal{F}_{\boldsymbol{k}%
}^{z}$ and non-adiabatic effects embedded in $\bar{\eta}_{n}^{i}$'s. In the
following, we illustrate these points in $AB$-stacked bilayer graphenes.

\section{The band manifolds of $AB$-stacked bilayer graphene}

The energy dispersion relation, including both active and remote manifolds,
for $AB$-stacked bilayer graphene can be analytically obtained. In the basis
of the sublattices of the two layers 1 and 2, labeled as $%
A_{1},B_{1},A_{2},B_{2}$ (see Fig. \ref{fig1}(a)), the effective four-band
Hamiltonian near the Dirac point $K\left( \text{or }K^{\prime }\right) $ can
be expressed as \cite{Mc2}
\end{subequations}
\begin{equation}
\mathcal{H}\left( \boldsymbol{k}\right) =\left(
\begin{array}{cccc}
\epsilon _{A_{1}} & f\pi ^{\dagger } & -f_{4}\pi ^{\dagger } & f_{3}\pi \\
f\pi & \epsilon _{B_{1}} & \gamma _{1} & -f_{4}\pi ^{\dagger } \\
-f_{4}\pi & \gamma _{1} & \epsilon _{A_{2}} & f\pi ^{\dagger } \\
f_{3}\pi ^{\dagger } & -f_{4}\pi & f\pi & \epsilon _{B_{2}}%
\end{array}%
\right)  \label{H}
\end{equation}%
with $\pi =\xi k_{x}+ik_{y}$, $f=\sqrt{3}a\gamma _{0}/2$, and $f_{3,4}=\sqrt{%
3}a\gamma _{3,4}/2$ in which $\xi =\pm 1$ is the valley index and $a$ is the
lattice constant. The on-site potentials can be represented explicitly as $%
\epsilon _{A_{1}}=-\frac{1}{2}U$, $\epsilon _{B_{1}}=\frac{1}{2}\left(
-U+2\Delta ^{\prime }\right) $, $\epsilon _{A_{2}}=\frac{1}{2}\left(
U+2\Delta ^{\prime }\right) $, and $\epsilon _{B_{2}}=\frac{1}{2}\left(
U-\delta _{AB}\right) $, with $U$ the interlayer asymmetry between the two
layers, $\Delta ^{\prime }$ for an energy difference between dimer and
non-dimer sites. The intralayer hopping between the $A$ and $B$ sites is
given by $\gamma _{0}$ and the meanings of various interlayer couplings $%
\gamma _{i}$ with $i=1,3,4$ are indicated in Fig.~\ref{fig1}(a). Following
Ref.~\cite{Kuz}, we fix $a=2.46\mathrm{\mathring{A}}$, $\gamma _{0}=3.16%
\mathrm{eV}$, $\gamma _{1}=0.381\mathrm{eV}$ and $\gamma _{4}=\Delta
^{\prime }=0$ throughout this work.

The band energies obtained from diagonalising Eq.~(\ref{H}) is illustrated
in Fig.~\ref{fig1}(b) clearly show two bands around the Fermi energy. They
are the lowest conduction band and the highest valence band, indexed by $c$
and $v$ respectively, that form the active manifold. The remote bands
consist of the higher conduction band $c_{R}$ and the lower valence band $%
v_{R}$ (see Fig.~\ref{fig1}(b)). The valley Hall current is $\boldsymbol{J}^{%
\text{H}}=\boldsymbol{J}_{A}^{\text{H}}+\boldsymbol{J}_{R}^{\text{H}}$
detailed in Eq.~(\ref{crnt-dH}). By Eqs.~(\ref{crnt-dH-2}) and (\ref%
{decVelo-R}), we see $\boldsymbol{J}_{R}^{\text{H}}$ is perpendicular to $%
\boldsymbol{E}$. As shown in an earlier study that for active manifolds with
two bands, $\boldsymbol{J}_{A}^{\text{H}}$ is also perpendicular to $%
\boldsymbol{E}$ \cite{Tu1}. We apply the polar coordinate, i.e., $%
\boldsymbol{E}=E\hat{\rho}$ with the basis vector $\hat{\rho}=\hat{x}\cos
\theta +\hat{y}\sin \theta $ and $\hat{\theta}=-\hat{x}\sin \theta +\hat{y}%
\cos \theta $. Here $E\equiv \left\vert \boldsymbol{E}\right\vert $ and $%
\theta $ is the angle between $\boldsymbol{E}$ and $\hat{x}$. It is then
sufficient to investigate the scalars $J_{A/R}^{\text{H}}=\hat{\theta}\cdot
\boldsymbol{J}_{A/R}^{\text{H}}$ and $J_{\text{occ}/\text{coh},R}^{\text{H}}=%
\hat{\theta}\cdot \boldsymbol{J}_{\text{occ}/\text{coh},R}^{\text{H}}$. For
definiteness, we place the chemical potential $\mu $ in the middle of the
gap. Then only the lower-energy hybridised band, indexed by $i_{v}$, needs
to be counted at very low temperatures in $\sum_{i}$ in Eq.~(\ref{crnt-dH-2}%
). This simplifies $J_{\text{occ}/\text{coh},R}^{\text{H}}=-e\int_{%
\boldsymbol{k}}\bar{v}_{i_{v}}^{\text{occ}/\text{coh},R}$ with $\bar{v}%
_{i_{v}}^{\text{occ}/\text{coh},R}=\hat{\theta}\cdot \bar{\boldsymbol{v}}%
_{i_{v}}^{\text{occ}/\text{coh},R}$ (see Eqs.~(\ref{decVelo-R}) and (\ref%
{crnt-dH})) and $J_{A}^{\text{H}}=-e\int_{\boldsymbol{k}}\bar{v}_{i_{v}}^{%
\text{coh},A}$ with $\bar{v}_{i_{v}}^{\text{coh},A}=\hat{\theta}\cdot \bar{%
\boldsymbol{v}}_{i_{v}}^{\text{coh},A}$ (see Eqs.~(\ref{crnt-A}) and (\ref%
{crnt-dH})).

\section{Intra- versus inter-manifold contributions to the Hall currents}

Since $\boldsymbol{E}=E\hat{\rho}(\theta )$ is a vector, $J_{A/R}^{\text{H}}$
in principle depends on both $E$ and $\theta $. When the dispersion of the
system is isotropic, then the dependence of $J^{\text{H}}$ on $\theta $
disappears. The two contributions $J_{A}^{\text{H}}$ and $J_{R}^{\text{H}}$
should depend on $E$ very distinctively. When the dispersion is anisotropic,
then $J_{A}^{\text{H}}$ and $J_{R}^{\text{H}}$ should also depend on $\theta
$, which characterises the orientation of the applied field, in very
different manners. We now discuss explicitly how these are revealed.

\begin{figure}[tbp]
\begin{center}
\includegraphics[width=0.48\textwidth]{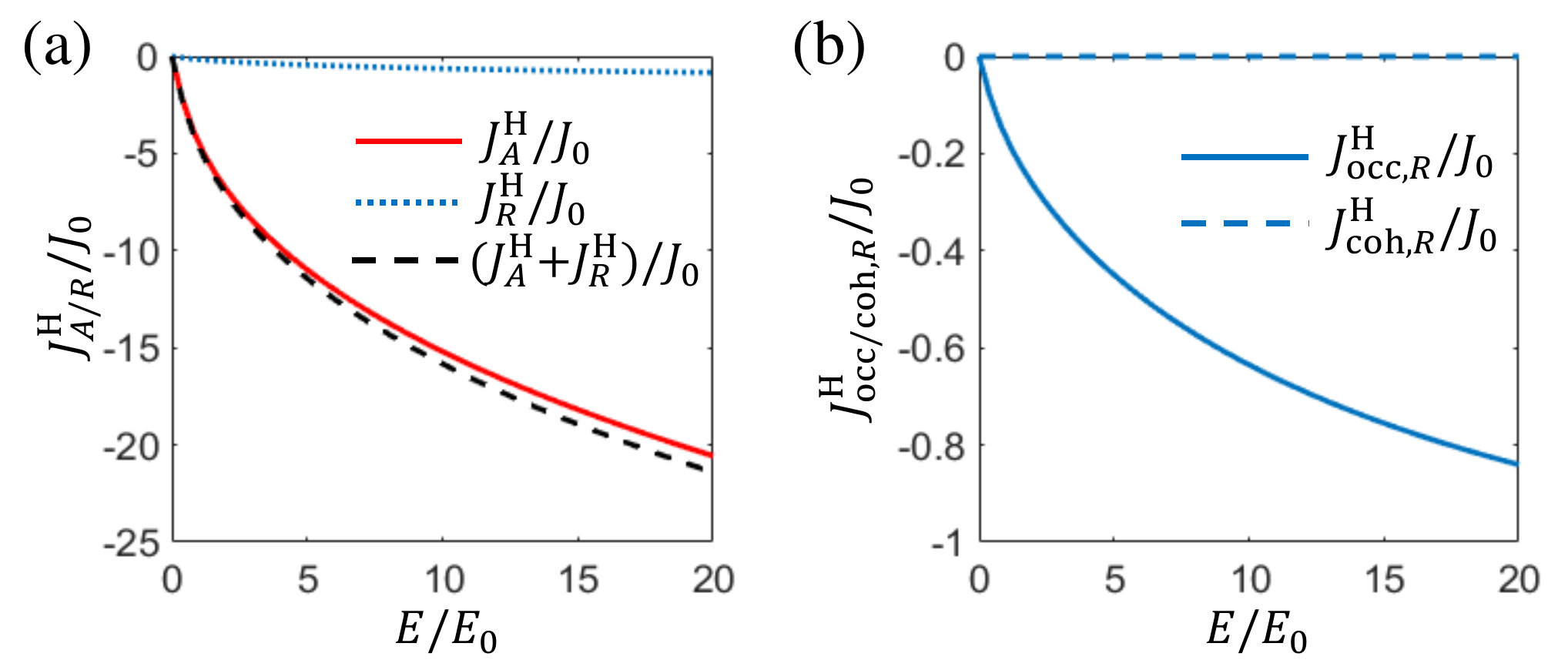}
\end{center}
\caption{(Color online) Under isotropic dispersion with $\protect\gamma %
_{3}=0$, different contributions to the Hall currents only depend on the
magnitude $E$ of applied field $\boldsymbol{E}$. (a): The non-adiabatic
intra-active-manifold contribution $J_{A}^{\text{H}}$ and adiabatic
inter-manifold contribution $J_{R}^{\text{H}}$ show significant quantitative
difference. (b): The two components $J_{\text{occ},R}^{\text{H}}$ and $J_{%
\text{coh},R}^{\text{H}}$ in Eq.~(\protect\ref{remote1}). Under such
isotropic dispersion, only the diagonals of the non-Abelian matrix are
involved in the nonzero contribution $J_{\text{occ},R}^{\text{H}}$ to $%
J_{R}^{\text{H}}$. Here the currents are in unit of $J_{0}=e^{2}E_{0}/h$,
where $E_{0}=1.15\mathrm{V/\protect\mu m}$ is chosen such that MAX $r_{c,v}\left(
\boldsymbol{k},E_{0}\right)=1$ (see Eq. (\protect\ref{r})). The energy gap
is $U=0.01\mathrm{eV}$.}
\label{fig2}
\end{figure}

The isotropy of dispersion here for the $AB$-stacked bilayer graphenes is
determined by $\gamma _{3}$, an interlayer skew hopping parameter (see Fig.~%
\ref{fig1}(a)). With $\gamma _{3}=0$, the distinction between $J_{A}^{\text{H%
}}$ and $J_{R}^{\text{H}}$ is simply displayed in the order-of-magnitude
difference between the current values, plotted as functions of $E$ in Fig.~%
\ref{fig2}(a) in unit of $E_{0}$ \cite{rr}. By defining
\begin{equation}
r_{n,m}\left( \boldsymbol{k},\boldsymbol{E}\right) =\left\vert\frac{\left[ \bar{\mathcal{R}}%
_{\boldsymbol{k}}\right] _{n,m}\cdot e\boldsymbol{E}}{\varepsilon
_{n}-\varepsilon _{m}}\right\vert  \label{r}
\end{equation}%
as a dimensionless quantity that measures the ratio between the
field-induced interband coupling to the gap size \cite{Tu1} at a given $%
\boldsymbol{k}$, $E_{0}$ is the field strength that satisfies MAX $r_{c,v}\left(
\boldsymbol{k},E_{0}\right) =1$. One can calculate $J^{\text{H}}$ by the
full-band formulation discussed in Ref. \cite{Tu}, without separating the
contributions into $J_{A}^{\text{H}}$ and $J_{R}^{\text{H}}$. We have
verified that the computations here based on Eq. (\ref{crnt-dH}) agree with
the results of full-band calculation. The consistency with full-band
calculation signifies that it indeed makes sense to simultaneously speak of
the adiabatic contribution (from coupling to remote band), and the
non-adiabatic contribution (from intra-manifold coupling) to the Hall effect
over a sizable range of the electric field. In addition, under the condition
$\gamma _{3}=0$, we found that $J_{R}^{\text{H}}$ is entirely dominated by $%
J_{\text{occ},R}^{\text{H}}$ with vanishing contribution from $J_{\text{coh}%
,R}^{\text{H}}$ (see Fig.~\ref{fig2}b)). We will show that when $\gamma
_{3}\neq 0$, more interesting differences between $J_{\text{occ},R}^{\text{H}%
}$ and $J_{\text{coh},R}^{\text{H}}$ can arise.

\begin{figure}[tbp]
\begin{center}
\includegraphics[width=0.48\textwidth]{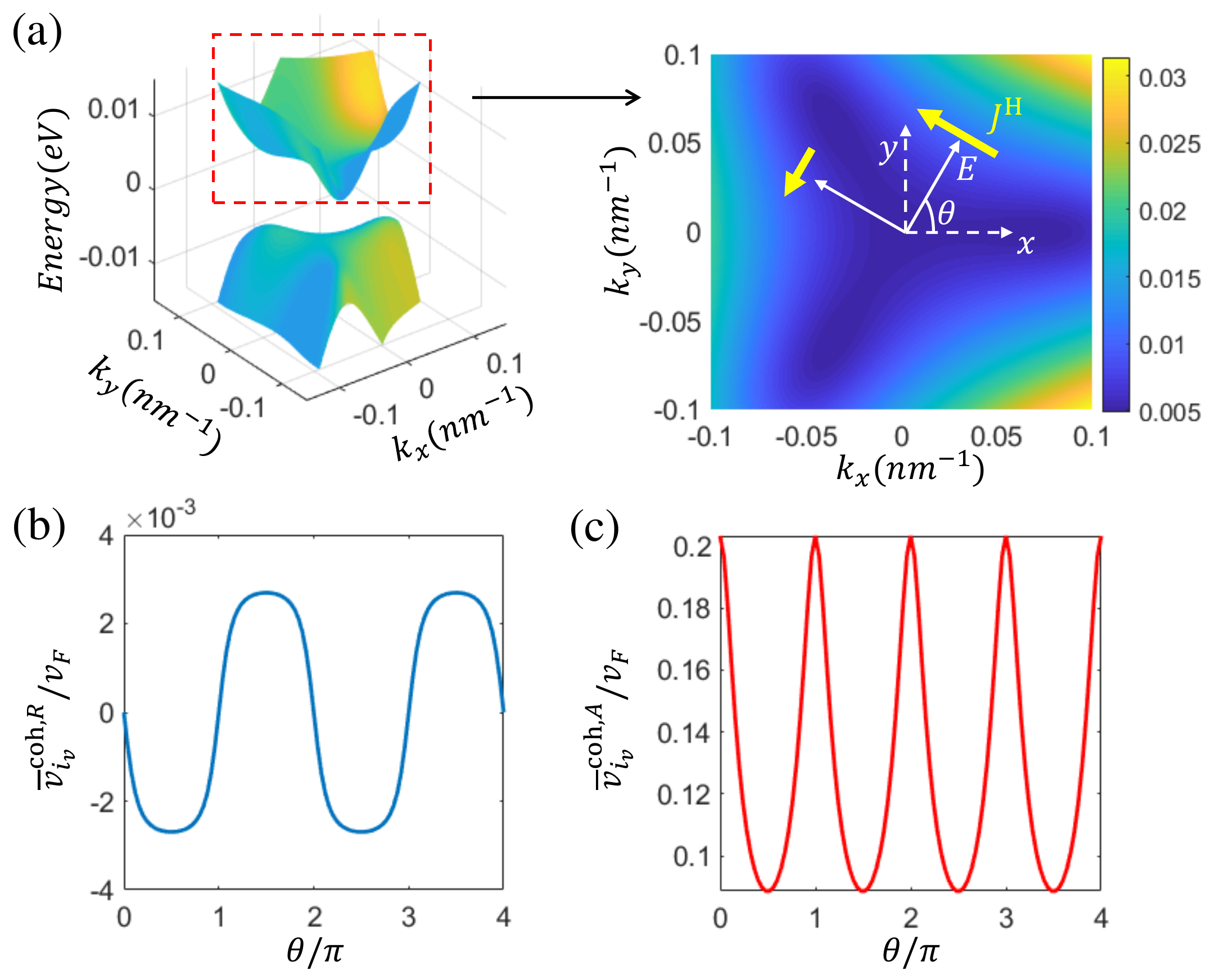}
\end{center}
\caption{(Color online) (a): The anisotropic dispersion of the bands near
the fermi energy, demonstrated with $\protect\gamma _{3}=0.38\mathrm{eV}$
\protect\cite{Kuz}. Inset: The trigonal warping of active conduction band,
implying that the magnitude of the Hall current (yellow arrows of different
lengths) depends on the angle $\protect\theta $ between $\boldsymbol{E}$
(solid white arrows) and the crystal axis, here taken as $\hat{x}$ (dashed
white arrow). (b) and (c): Different $\protect\theta $-dependencies of the
anomalous velocities $\bar{v}_{i_{v}}^{\text{coh},R}$ and $\bar{v}_{i_{v}}^{%
\text{coh},A}$ respectively, giving rise to distinct periods in $\protect%
\theta $ and determining the anisotropic responses of $J_{R}^{\text{H}}$ and
$J_{A}^{\text{H}}$ to $\protect\theta $ (see the main text for details). The
specific choice of $\boldsymbol{k}$ is irrelevant. Here we use $k_{x}=0.01$%
\textrm{\AA } and $k_{y}=0$. Other parameters follow that used in Fig.~%
\protect\ref{fig2} with $E=E_{0}$ here and $v_{F}=\protect\sqrt{3}a\protect%
\gamma _{0}/\hbar $ as the Fermi velocity.}
\label{fig3}
\end{figure}

When $\gamma _{3}\neq 0$ the anisotropy of the dispersion appears (see Fig. %
\ref{fig3}(a)), allowing non-trivial $\theta$-dependence of $J_{R}^{\text{H}%
}=J_{\text{occ},R}^{\text{H}}+J_{\text{coh},R}^{\text{H}}$ and $J_{A}^{\text{%
H}}$. From Eqs.~(\ref{remote-H-decomp}) and (\ref{decVelo-R}), we see on one
hand, $J_{\text{occ},R}^{\text{H}}$ depends on $\theta$ through the band
occupations $\left\vert \bar{\eta}_{c}^{i_{v}}\right\vert ^{2}$ and $%
\left\vert \bar{\eta}_{v}^{i_{v}}\right\vert^{2}$ that have the same period
in $\theta$ in $\bar{v}_{i_{v}}^{\text{occ},R}$. On the other hand, $J_{%
\text{coh},R}^{\text{H}}$ gains $\theta $-dependence through the interband
coherence $\left( \bar{\eta}_{c}^{i_{v}}\right)^{\ast }\bar{\eta}%
_{v}^{i_{v}} $ in $\bar{v}_{i_{v}}^{\text{coh},R}$. Explicitly, we have
\begin{equation}
\left\vert \bar{\eta}_{c/v}^{i_{v}}\right\vert ^{2}=\frac{1\mp \varepsilon
_{g}/\bar{\varepsilon}_{g}}{2},~\left( \bar{\eta}_{c}^{i_{v}}\right) ^{\ast }%
\bar{\eta}_{v}^{i_{v}}=-\frac{\left[ \bar{\mathcal{R}}_{\boldsymbol{k}}%
\right] _{v,c}\cdot \boldsymbol{E}}{\bar{\varepsilon}_{g}}  \label{bamp-2x2}
\end{equation}%
where $\varepsilon _{g}=\varepsilon _{c}-\varepsilon _{v}$ and
\begin{equation}
\bar{\varepsilon}_{g}=\sqrt{\varepsilon _{g}^{2}+4\left\vert \left[ \bar{%
\mathcal{R}}_{\boldsymbol{k}}\right] _{v,c}\cdot \boldsymbol{E}\right\vert
^{2}}  \label{renorm-gap-0}
\end{equation}%
is the renormalized gap. Using $\boldsymbol{E}=E(\hat{x}\cos \theta +\hat{y}%
\sin \theta )$, we straightforwardly see that $\left( \bar{\eta}%
_{c}^{i_{v}}\right) ^{\ast }\bar{\eta}_{v}^{i_{v}}$ has a $\theta $-period
that is twice of that of $\left\vert \bar{\eta}_{c}^{i_{v}}\right\vert ^{2}$
and $\left\vert \bar{\eta}_{v}^{i_{v}}\right\vert ^{2}$. So the $\theta $%
-period of $J_{R}^{\text{H}}$ is led by that of $J_{\text{coh},R}^{\text{H}}$%
, which is associated with the interband coherence that appears in $\bar{v}%
_{i_{v}}^{\text{coh},R}$ of Eq.~(\ref{decVelo-R}).

We now turn to the $\theta $-dependence in $J_{A}^{\text{H}}$. From Eqs.~(%
\ref{crnt-dH}) and (\ref{crnt-A}), we find that $J_{A}^{\text{H}}$ depends
on $\theta $ through the term $\left( \bar{\eta}_{c}^{i_{v}}\right) ^{\ast }%
\bar{\eta}_{v}^{i_{v}}\hat{\theta}\cdot \partial \mathcal{H}_{A}/\partial
\hbar \boldsymbol{k}$ that appears in $\bar{v}_{i_{v}}^{\text{coh},A}$. Here
$\hat{\theta}\cdot \partial \mathcal{H}_{A}/\partial \hbar \boldsymbol{k}%
=\left( -\sin \theta \partial \mathcal{H}_{A}/\partial \hbar {k}_{x}+\cos
\theta \partial \mathcal{H}_{A}/\partial \hbar {k}_{y}\right) $ which makes
the $\theta $-period of $\left( \bar{\eta}_{c}^{i_{v}}\right) ^{\ast }\bar{%
\eta}_{v}^{i_{v}}\hat{\theta}\cdot \partial \mathcal{H}_{A}/\partial \hbar
\boldsymbol{k}$ only half as that of $\left( \bar{\eta}_{c}^{i_{v}}\right)
^{\ast }\bar{\eta}_{v}^{i_{v}}$. These two different $\theta $-dependencies,
namely, $\left( \bar{\eta}_{c}^{i_{v}}\right) ^{\ast }\bar{\eta}_{v}^{i_{v}}$
and $\left( \bar{\eta}_{c}^{i_{v}}\right) ^{\ast }\bar{\eta}_{v}^{i_{v}}\hat{%
\theta}\cdot \partial \mathcal{H}_{A}/\partial \hbar \boldsymbol{k}$, are
revealed correspondingly in the velocities $\bar{v}_{i_{v}}^{\text{coh},R}$
and $\bar{v}_{i_{v}}^{\text{coh},A}$, plotted as functions of $\theta $ in
Fig.~\ref{fig3}(b) and (c) respectively. It clearly shows the distinctively
different periods.

%Physically, $J_{A}^{\text{H}}$ arises from direct coupling between the active bands (see Fig.~\ref{fig1}(c)), where the interband coherence appears in the current with the involvement of the velocity matrix $\partial \mathcal{H}_{A}/\partial \hbar \boldsymbol{k}$ evaluated between these active bands (interpreted from Eqs.~(\ref{crnt-dH}) and (\ref{crnt-A})). By the same token, $J_{\text{coh},R}^{\text{H}}$ arises from coupling of the active bands to the remote bands and should involve the velocity matrix elements between remote and active bands.

%which indirectly contributes to the coherence between the two active bands (see Fig.~\ref{fig1}(d)). The velocity matrix evaluated between remote and active band is replaced by the induced Berry curvatures.

%We plot the interband coherence $\left(\bar{\eta}_{c}^{i_{v}}\right)^{\ast
%}\bar{\eta}_{v}^{i_{v}}$ as a function of $\theta$ in Fig.~\ref{fig3}(b).

From the above analysis with also the aids of Eqs.~(\ref{decVelo-R}) and (%
\ref{crnt-A}), the $\theta $-period of the average carrier velocity $\bar{v}%
_{i_{v}}^{\text{coh},R}$ is twice as that of $\bar{v}_{i_{v}}^{\text{coh},A}$
at arbitrary $\boldsymbol{k}$ regardless of the dispersion is anisotropic or
isotropic. The currents are obtained by integrating the average carrier
velocity over $\boldsymbol{k}$. The above discussed $\theta $-period
difference between $\bar{v}_{i_{v}}^{\text{coh},R}$ and $\bar{v}_{i_{v}}^{%
\text{coh},A}$ appears also in the difference between $J_{R}^{\text{H}}$ and
$J_{A}^{\text{H}}$ only when the dispersion of the system is anisotropic.
With $\gamma _{3}\neq 0$, which leads to a well-known trigonal-warped
dispersion \cite{Mc2,Kech}, we show in Fig.~\ref{fig4}(a) at a given valley $%
\xi =1$ both current contributions $J_{R}^{\text{H}}$ and $J_{A}^{\text{H}}$%
. It clearly shows that while $J_{R}^{\text{H}}$ follows the $C_{3}$
symmetry of the energy dispersion, $J_{A}^{\text{H}}$ exhibits $C_{6}$
symmetry. The $\theta $-period of the current is different from the $%
\boldsymbol{k}$-resolved average carrier velocity due to integration over $%
\boldsymbol{k}$, which manifests the symmetry of the dispersion.
Nevertheless, the character that $\bar{v}_{i_{v}}^{\text{coh},R}$ has its $%
\theta $-period twice as that of $\bar{v}_{i_{v}}^{\text{coh},A}$ is fully
revealed in the physical observable that $J_{R}^{\text{H}}$'s $\theta $%
-period is twice of $J_{A}^{\text{H}}$'s. The intuitively anticipated
qualitative difference between $J_{A}^{\text{H}}$ and $J_{R}^{\text{H}}$
from Fig.~\ref{fig1}(c) and (d) is thus concretely illustrated through
different periods in $\theta $.

%In short, the non-adiabatic intra-active-manifold contribution $J^{\text{H}}_{A}$ is on one hand quantitatively different from the adiabatic inter-manifold contribution $J^{\text{H}}_{R}$ by an order of magnitude, due to the perturbation role of the applied in inter-manifold transitions. On the other hand, these two contributions also differ qualitatively

%the differential conductivities, defined by,
%\begin{equation}
%\sigma_{\rho }^{A/R}=\frac{\partial {J}_{A/R}^{\text{H}}}{\partial {E}}
%,~\sigma _{\theta }^{A/R}=\frac{1}{E}\frac{\partial {J}_{A/R}^{\text{H}}}{\partial \theta }.
%\label{diffcndv-def}
%\end{equation}
%as a function of $\theta$ at a finite $E\ne0$.

\section{Two faces of multiple-band effects: non-adiabatic and non-Abelian}

\subsection{revelation on the Hall currents}

We now focus on the remote band contribution $J_{R}^{\text{H}}$ to reveal
the relation between the underlying non-adiabatic dynamics and the
non-Abelian characters raised from the multi-band effects in the active
manifold. As we have mentioned before, when the active bands are fully
degenerate and occupied, the linear conductivity is given by $\sigma ^{\text{%
H}}=\left. \partial J^{\text{H}}/\partial {E}\right\vert _{E=0}=\left(
e^{2}/\hbar \right) \int_{\boldsymbol{k}}\text{Tr}\left( \mathcal{F}_{%
\boldsymbol{k}}^{z}\right) $, which depends neither on $E$ nor on $\theta $,
as a consequence of its definition. Furthermore, this linear conductivity
only involves the band diagonals of the Berry curvature matrix $\mathcal{F}_{%
\boldsymbol{k}}^{z}$.

\begin{figure}[tbp]
\begin{center}
\includegraphics[width=0.48\textwidth]{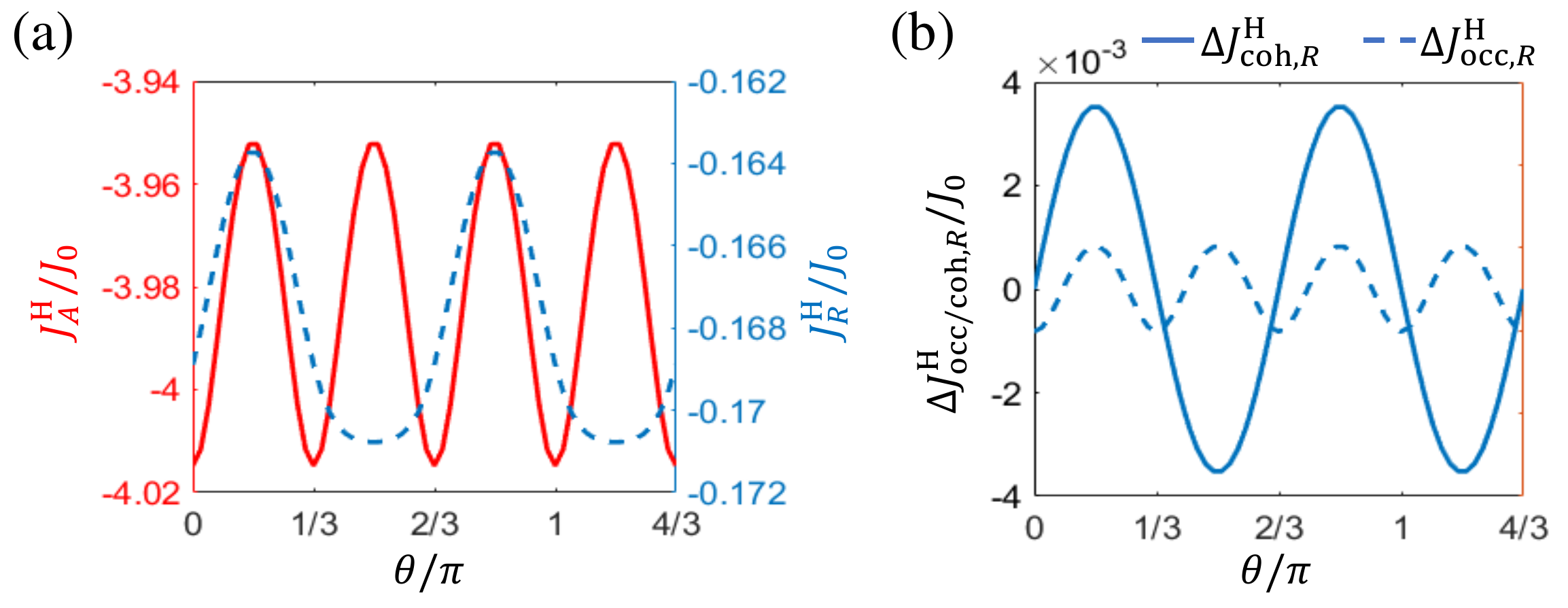}
\end{center}
\caption{(Color online) (a): The qualitative difference between $J_{R}^{%
\text{H}}$ and $J_{A}^{\text{H}}$ revealed through distinct symmetries ($%
C_{3}$ for $J_{R}^{\text{H}}$ and $C_{6}$ for $J_{A}^{\text{H}}$). (b):
Angular variations of inter-band coherence ($\Delta {J}_{\text{coh},R}^{%
\text{H}}$ as the solid line) and band-occupation ($\Delta {J}_{\text{occ}%
,R}^{\text{H}}$ as the dashed line) contributions, respectively to the
remote-band-mediated Hall current $J_{R}^{\text{H}}$, contrasting the
diagonal and the off-diagonal involvements of the non-Abelian Berry
curvature matrices. The parameters are same as what we take in Fig. \protect
\ref{fig3} for both (b) and (c). The angular averages are $\bar{J}_{\text{occ%
},R}^{\text{H}}=0.168J_{0}$ and $\bar{J}_{\text{coh},R}^{\text{H}}=0$.}
\label{fig4}
\end{figure}

%Here we show that within $J_{R}^{\text{H}}=$, we have  $J_{\text{occ},R}^{\text{H}}\ll J_{\text{coh},R}^{\text{H}}$.
As discussed before, anisotropy of the dispersion can be revealed in the
Hall currents under the finite electric fields. In order to extract the $%
\theta $-dependence more exclusively, we define $\Delta {J}_{\text{coh}/%
\text{occ},R}^{\text{H}}=J_{\text{coh}/\text{occ},R}^{\text{H}}-\bar{J}_{%
\text{coh}/\text{occ},R}^{\text{H}}$, where we subtract the angular average $%
\bar{J}_{\text{coh}/\text{occ},R}^{\text{H}}=\left( 1/\theta _{T}^{\text{coh}%
/\text{occ},R}\right) \int_{0}^{\theta _{T}^{\text{coh}/\text{occ},R}}\text{d%
}\theta {J}_{\text{coh}/\text{occ},R}^{\text{H}}$ in which $\theta _{T}^{%
\text{coh}/\text{occ},R}$ denotes the $\theta $-period of $J_{\text{coh}/%
\text{occ},R}^{\text{H}}$. We plot $\Delta {J}_{\text{coh},R}^{\text{H}}$
(solid line) and $\Delta {J}_{\text{occ},R}^{\text{H}}$ (dashed line) as a
function of $\theta $ in Fig.~\ref{fig4}(b), which shows that $\theta _{T}^{%
\text{coh},R}=2\theta _{T}^{\text{occ},R}$, as expected due to the reasons
already discussed for Fig.~\ref{fig3}. More interestingly, in Fig.~\ref{fig4}%
(b) we see that $\Delta {J}_{\text{coh},R}^{\text{H}}$ displays a variation
with respect to $\theta $ that is clearly greater than that of $\Delta {J}_{%
\text{occ},R}^{\text{H}}$. This is because the interband coherence $\left(
\bar{\eta}_{c}^{i_{v}}\right) ^{\ast }\bar{\eta}_{v}^{i_{v}}$ in general is
more sensitive to $\theta $ than the occupation $\left\vert \bar{\eta}%
_{c/v}^{i_{v}}\right\vert ^{2}$ since the former is one order larger in $%
E\cos \theta $ and $E\sin \theta $ than the latter (see Eq.~(\ref{bamp-2x2}%
)). Note that in terms of $J_{\text{coh}/\text{occ},R}^{\text{H}}=\Delta {J}_{\text{coh}/\text{occ},R}^{\text{H}}+\bar{J}_{\text{coh}/\text{occ},R}^{\text{H}}$, we still have $J_{\text{occ}, R}^{H}\gg J_{\text{coh},R}^{H}$ for $\gamma_{3}\ne0$. This is consistent with the observation in Fig.~\ref{fig2}(b) where $\gamma_{3}=0$. Importantly, it is the anisotropy (sensitivity to $\theta$'s variation due to $\gamma_{3}\ne0$) that pronounces the role of the off-diagonal term, due to which the Berry curvature is said to be non-Abelian. The underlying non-adiabatic dynamics, characterised by $\theta$-dependence at finite $\boldsymbol{E}$, thus manifests the non-Abelian characters of $\mathcal{F}_{\boldsymbol{k}}^{z}$, differentiating between
the diagonal $J_{\text{occ},R}^{\text{H}}$ and off-diagonal $J_{\text{coh},R}^{\text{H}}$ contributions.

%the latter depends on $\theta$ only through $1/\bar{\varepsilon}_{g}$ while the former is additionally modulated by $\theta$ one order higher in $E$ by $\left[\bar{\mathcal{R}}_{%
%\boldsymbol{k}}\right]_{v,c}\cdot\hat{\rho}\left(\theta\right)$.

%Since angular dependence is acquired through the non-perturbation effects of $\boldsymbol{E}$ on the band amplitudes $\bar{\eta}_{n}$'s (see Eqs.~(\ref{decVelo-R}),(\ref{crnt-dH}),(\ref{remote-H-decomp})),

%Note that perturbation inspection of the anomalous velocity shows that the contributions from
%$\left(\bar{\eta}_{c}^{i}\right)^{\ast}\bar{\eta}_{v}^{i}$ appears readily on the first order of $E$ while the lowest nonzero order of $E$ from the contributions of
%$\left\vert\bar{\eta}_{c}^{i}\right\vert^{2}$ and $\left\vert\bar{\eta}_{v}^{i}\right\vert^{2}$

\subsection{microscopic picture from $\boldsymbol{k}$-resolved processes}

So far the two faces of multiple-band effects have been investigated through the $\boldsymbol{k}$-integrated currents. To have a deeper insight, we revisit the $\boldsymbol{k}$-resolved dimensionless quantity, Eq.~(\ref{r}). On one hand, with the band indices $n=c,m=v\in{A}$, Eq.~(\ref{r}) measures the strength of non-adiabatic intra-manifold transition with respect to the gap between active bands $c$ and $v$. On the other hand, the underlying constituents of the non-Abelian Berry curvature Eq.~(\ref{NAB}), which can be rewritten as $\left[\mathcal{F}_{\boldsymbol{k}}^{\alpha \beta}\right]_{\left(m^{\prime},m\right)\in{A}}=\sum_{n\in{R}}\left\{i\left[\mathcal{R}_{k_{\alpha}}\right]_{m^{\prime},n}
\left[\mathcal{R}_{k_{\beta}}\right]_{n,m}
-(k_{\alpha}\leftrightarrow k_{\beta})\right\}$, are the inter-manifold Berry connections $\left[\mathcal{R}_{k_{\beta}}\right]_{n,m}$ contained in Eq.~(\ref{r}) with $n\in{R}$ and $m\in{A}$. We thus are able to explore the multiple-band effects from the $\boldsymbol{k}$-resolved information of Eq.~(\ref{r}).

\begin{figure}[tbp]
\begin{center}
\includegraphics[width=0.48\textwidth]{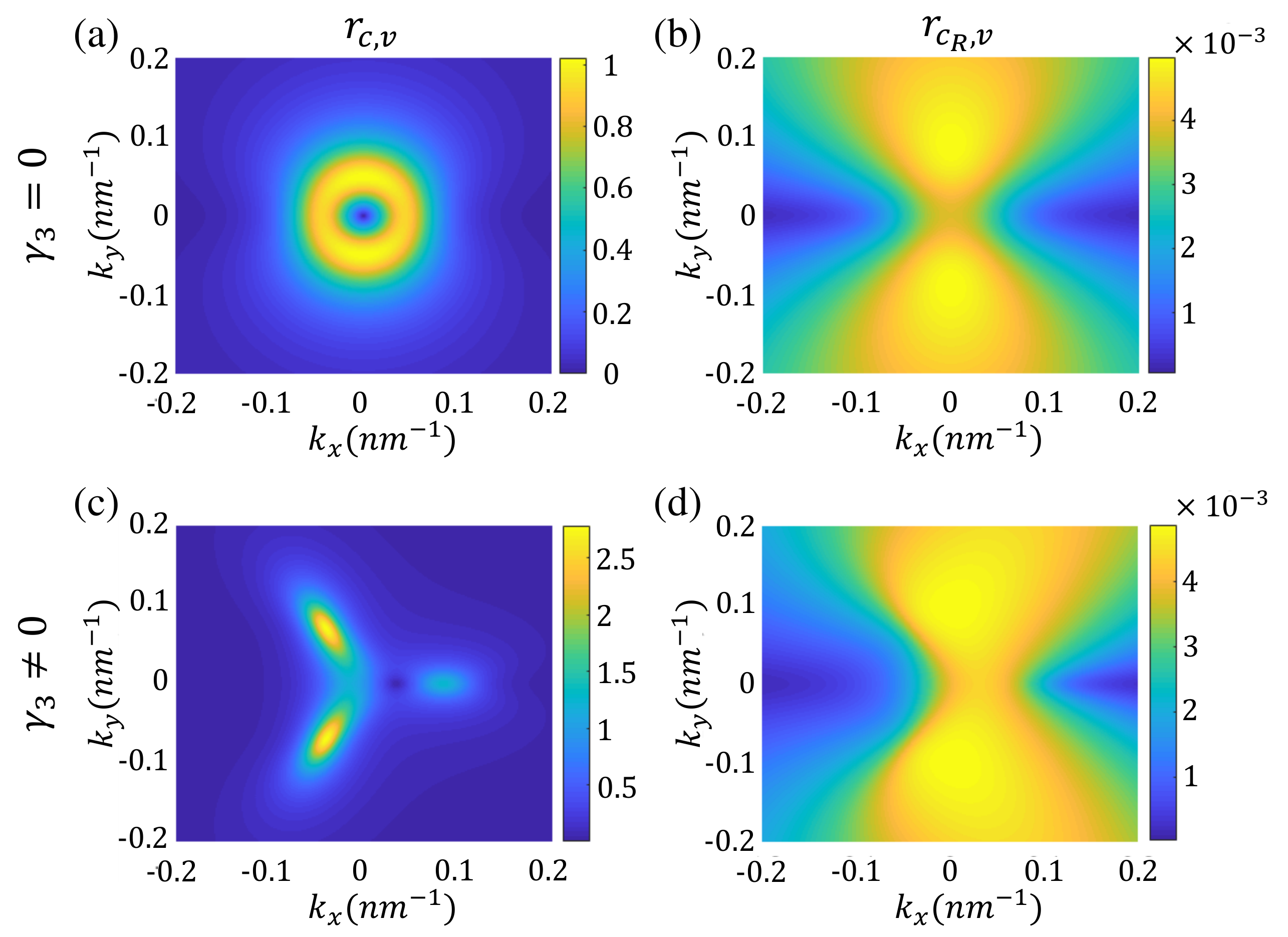}
\end{center}
\caption{(Color online) Visualization of the dimensionless non-adiabatic measure $
r_{n,m}(\boldsymbol{k},\boldsymbol{E}=E_{0}\hat{x})$ in the $\boldsymbol{k}$-space for both isotropic ($\gamma_{3}=0$ as (a) and (b)) and
anisotropic ($\protect\gamma _{3}=0.38\mathrm{eV}$ as (c) and (d)) dispersions. Other parameters are the same as
in Fig.~\protect\ref{fig2} (for $\gamma_{3}=0$) and Fig.~\ref{fig3} (for $\gamma_{3}\ne0$). Similarity between (b) and (d) is contrasted with dissimilarity between (a) and (c) to reveal the distinctness of non-adiabatic interband transitions as a multiple-band effect.}
\label{fig5}
\end{figure}

In Fig.~\ref{fig5}, we inspect $\boldsymbol{k}$-dependencies of $r_{c,v}\left(\boldsymbol{k},\boldsymbol{E}\right)$ (for non-adiabatic intra-active-manifold dynamics) and $r_{c_{R},v}\left(\boldsymbol{k},\boldsymbol{E}\right)$ (for constituents of non-Abelian Berry curvatures) respectively for both cases $\gamma_{3}=0$ and $\gamma_{3}\ne0$.
The $\boldsymbol{k}$-dependence of $r_{c,v}$ follows the symmetry of $\boldsymbol{k}$-dependence of band energies $\varepsilon_{c/v}(\boldsymbol{k})$ (see Fig.~\ref{fig5}(a) and (c) with respective energy dispersion in mind).\cite{note-2} In contrast, $r_{c_{R},v}$ for $\gamma_{3}=0$ and that for $\gamma_{3}\ne0$ have similar patterns in $\boldsymbol{k}$-dependence (see Fig.~\ref{fig5}(b) and (d)) that are distinct from the $\boldsymbol{k}$-dependence pattern of $r_{c,v}$'s in Fig.~\ref{fig5}(a) and (c). Regardless of $\gamma_{3}=0$ or $\gamma_{3}\ne0$, $r_{c_{R},v}$'s do not follow the symmetry of energy dispersion in $\boldsymbol{k}$-dependence.
This provides the insight that the non-adiabatic interband transitions and the transitions giving rise to non-Abelian Berry curvatures, where both are responsible for Hall currents, are fundamentally different in their natures.

As discussed before with an analysis of $J^{\text{H}}$'s dependence on $\boldsymbol{E}$'s orientation, such qualitative difference can be seen from $\boldsymbol{k}$-integrated currents only when the energy dispersion is anisotropic. Crucially, even without energy anisotropy, the qualitative differences in the microscopic $\boldsymbol{k}$-resolved processes already underlie the origin that the non-adiabatic contribution to the Hall current is physically distinct from the adiabatic one, as a result of multiple-band effects.

%for non-adiabatic transitions and the transitions giving rise to non-Abelian Berry curvatures

%Therefore, with the $\boldsymbol{k}$-resolved non-adiabatic measure Eq.~(\ref{r}), one is also to see not only the quantitative but more crucially also the qualitative difference between intra-manifold non-adiabatic dynamics and inter-manifold adiabatic transitions without relying on energy dispersion anisotropy and dependence on $\boldsymbol{E}$'s orientation as discussed before.

%The two faces of multiple-band effects thus can also be studied through investigating Eq.~(\ref{r}).

%In principle, the relative orientation between the vectors $\boldsymbol{E}$ and $\boldsymbol{k}$

%$r_{c_{R},v}\left(\boldsymbol{k},\boldsymbol{E}\right)$, which contains  between remote and active bands,

%$r_{c,v}\left(\boldsymbol{k},\boldsymbol{E}\right)$

\section{Conclusions and discussions}

In summary, through the studies of the valley Hall currents in typical $AB$%
-stacked bilayer graphenes as an example, we have demonstrated the
followings. (i): The adiabatic inter-manifold contribution and the
non-adiabatic intra-active-manifold contribution to the Hall current not
only differ quantitatively about an order of magnitude due to the
perturbation role of the electric field in the adiabatic inter-manifold
transition processes. They also differ qualitatively in terms of the
periodicity with respect to the angle of the applied electric field. (ii):
The non-Abelian characters due to the multiple-band nature of the active
manifold are manifested by the difference between the diagonal and the
off-diagonal contributions to the Hall current in terms not only of the
periodicity with respect to the angle of the applied electric field but also
of the angular variation amplitude.

The values of the angular periods and the shapes of the angular profiles for
different contributions to the Hall currents depend on the details of the
materials' electronic structures under consideration. Nevertheless, the
conclusions (i) and (ii), confirmed by the calculations specifically
designated to $AB$-stacked bilayer graphenes, are expected to be held also
for other materials with anisotropic energy dispersion such that the Hall
current at finite electric fields is sensitive to the orientation of the
field. This is because the general analysis leading to the above conclusions
is simply rooted to how band occupation and interband coherence differ
fundamentally in their responses to the applied electric field (as
exemplified for two-band active manifold in Eq.~(\ref{bamp-2x2})). This
fundamental difference also serves as the basis to manifest the
multiple-band induced non-Abelian effects in Hall currents. Interestingly,
this provides an intuitive approach toward the understanding of the
non-Abelian characters, totally based on physically observable carrier
transport currents, as complementary to the conventionally more
mathematically oriented understanding based on quantum geometry of the band
structures alone.

%The present attempt to identify the fundamental roles played by adiabatic and non-adiabatic interband transitions in Hall conduction also merits in that it offer a non-geometric approach to understand the non-Abelian characters of the gauge structures from multiple bands

\section*{Acknowledgments}

C. Li would like to thank D. W. Zhai and B. Fu for useful discussions. The
work is mainly supported by the Research Grants Council of Hong Kong
(HKU17306819 and AoE/P-701/20), and the University of Hong Kong (Seed
Funding for Strategic Interdisciplinary Research). M. W.-Y. Tu acknowledge
the the hospitality of Prof. Tay-Rong Chang with support from ministry of
technology and science in Taiwan of grand no. MOST110-2636-M-006-016. M.
W.-Y. Tu also acknowledge the hospitality of National Center for Theoretical
Science and thank Prof. Tse-Min Chen for useful discussions.

%\section*{References}


\begin{thebibliography}{99}
\bibitem{note-1} The ensemble-averaged velocity $\bar{\boldsymbol{v}}_{i}$
given in Eq. (\ref{eqvelo-x-i}) though looks similar to the velocity
expression of a single wave packet $\dot{\boldsymbol{r}}$ given by the first
line of Eq.~(\ref{eom-cc}) in terms of the decomposition into intra-manifold
and inter-manifold processes, they are fundamentally different. The velocity
in Eq.~(\ref{eom-cc}) is obtained by the full coherent evolution of a single
wave packet. Its value at a given instance on which the wave packet has been
driven to carry the momentum $\boldsymbol{k}$ is fully dependent on the
non-Markov history of band amplitudes and the Bloch momentum of the wave
packet. Henceforth $\dot{\boldsymbol{r}}$ is not fixed by the present $%
\boldsymbol{k}$ and $\boldsymbol{E}$ alone. The ensemble-averaged velocity
is obtained from taking into account the Markov decoherence process through
the evaluation of the density operator in the steady-state limit at each $%
\boldsymbol{k}$. Therefore, $\bar{\boldsymbol{v}}_{i}$ is fully determined
by $\boldsymbol{k}$ and $\boldsymbol{E}$. The integral in Eq.~(\ref{Bltzeq-4}%
) for the current arising from an ensemble of wave packets is then
well-defined. See more details in Ref.\cite{Tu}.
\bibitem{rr} We have shown analytically in Ref. \cite{Tu} that by taking the
dimensionless quantities $\left\vert \mathcal{\bar{R}}_{n_{R},m_{A}}\cdot
E/\left( \varepsilon _{n_{R}}-\varepsilon _{m_{A}}\right) \right\vert $,
where $n_{R}/m_{A}$ belong to remote/active bands, as perturbation to the
first order, $J_{R}^{\text{H}}$ reduces to the form with non-Abelian Berry
curvature as shown in Eqs. (\ref{decVelo-R}) and (\ref{crnt-dH}). Since this
is a perturbtion result, $J_{R}^{\text{H}}$ of course is expected to be
small. Here this anticipation is verified in the numerical calculation.
\bibitem{note-2} Usually $r_{c,v}$ is peaked at the Dirac point because the gap size there is the smallest. Nevertheless in Fig. \ref{fig5}(a), with $\gamma_{3}=0$ the Berry connection there vanishes, namely, $\left[\mathcal{R}_{\boldsymbol{k}=0}\right]_{c,v}=0$. The band structure is dramatically changed by $\gamma_{3}\ne0$ with the trigonal warping effect also visible in $r_{c,v}$. The directionality of $\boldsymbol{E}$ (here set in the direction of $\hat{x}$) slightly deviates $r_{c,v}$'s $\boldsymbol{k}$-dependence from the symmetry of energy dispersion, as a result of definition of Eq.~(\ref{r}). Noticeably, the anisotropy of energy dispersion with $\gamma_{3}\ne0$ is also accompanied by stronger non-adiabatic intra-manifold transition effects with respect to the same value of gap (see the differences in color bars between Fig. \ref{fig5}(a) and Fig. \ref{fig5}(c)).
\end{thebibliography}
\end{document}